
\input phyzzx

\def\dplus{=\hskip-5pt \raise 0.7pt\hbox{${}_\vert$} ^{\phantom 7}}
\def\dplusup{=\hskip-5.1pt \raise 5.4pt\hbox{${}_\vert$} ^{\phantom 7}}

\def\dplus{=\hskip-4.8pt \raise 0.7pt\hbox{${}_\vert$} ^{\phantom 7}}

\def\pmb#1{\setbox0=\hbox{#1} \kern-.025em\copy0\kern-\wd0
\kern0.05em\copy0\kern-\wd0 \kern-.025em\raise.0433em\box0}

\def\cM{{\cal M}}

\font\mybb=msbm10 at 12pt
\def\bb#1{\hbox{\mybb#1}}

\def\bE {\bb{E}}

\def\bM {\bb{M}}

\REF\pkta{P.K. Townsend, Phys. Lett. {\bf 350B} (1995) 184.} 
\REF\wita{E. Witten, Nucl. Phys. {\bf B443} (1995) 85.}
\REF\hs{G.T. Horowitz and A. Strominger, Nucl. Phys. {\bf B360}  (1991) 197.}
\REF\rrrbranes{M. J. Duff and J. X. Lu, Nucl. Phys. {\bf B416}(1994) 301.}
\REF\pol{J. Polchinski, Phys. Rev. Lett. {\bf 75} (1995) 4724.}
\REF\doug{M. Douglas, D. Kabat, P. Pouliot and S.H. Shenker, 
 Nucl. Phys. {\bf B485} (1997) 85, hep-th/9608024.}
\REF\gppkt{G. Papadopoulos and P.K. Townsend, 
Phys. Lett. {\bf B393} (1997)
59, hep-th/9609095.}
\REF\dhs{M. Dine, P. Huet and N. Seiberg, Nucl. Phys. {\bf B322} (1989) 301.}
\REF\dlp{J. Dai, R. Leigh and J. Polchinski, Mod. 
Phys. Lett.  {\bf A4}
(1989) 2073.}
\REF\douglas{M. Douglas, {\it Branes within Branes}, hep-th/9512077.}
\REF\polch{J. Polchinski, {\it TASI Lectures on D-branes},
hep-th/9611050.}
\REF\gg{M.B. Green and M. Gutperle, Nucl. Phys. {\bf B476} (1996) 484.}
\REF\sen{A. Sen, Phys. Rev. {\bf D54} (1996) 2964.}
\REF\gilad{G. Lifschytz, {\it Probing bound states of D-branes},
hep-th/9610125.}
\REF\memdy{J.M. Izquierdo, N.D. Lambert and 
G. Papadopoulos and  P.K.
Townsend, Nucl. Phys. {\bf B460}, (1996) 560.}
\REF\gp{G. Papadopoulos, {\it A brief guide to p-branes}, hep-th/9604068.}
\REF\glpt{M.B. Green, N.D. Lambert, 
G. Papadopoulos and P.K.  Townsend, Phys.
Lett. {\bf 384B} (1996) 86.}
\REF\tsey{ J. Russo and A. Tseytlin , Nucl. Phys. {\bf B490} (1997) 121,
hep-th/9611047.}
\REF\swh{J.H. Schwarz, Phys. Lett.{\bf B360} (1995) 13 
(E: {\bf B364} (1995) 252); Phys.
Lett. {\bf B367}  (1996) 97.}
\REF\wittenb{E. Witten, Nucl. Phys. {\bf B460} (1995) 335.}
\REF\bho{E. Bergshoeff, C.M. Hull 
and T. Ort\'\i n, Nucl. Phys.
{\bf B451} (1995) 547.}
\REF\pktb{P.K. Townsend, Phys. Lett. {\bf 227B} (1992) 285; 
{\it  Four
Lectures on M-theory}, hep-th/9612121. \break
E. Bergshoeff, L.A.J. London and P.K. Townsend,  
Class. Quantum   Grav. 
{\bf 9} (1992) 2545.}
\REF\duff{M J Duff and K S Stelle Phys. Lett. {\bf B253}(1991) 113.}
\REF\guv{R. G{\" u}ven, Phys. Lett. {\bf 276B} (1992) 49.}
\REF\gh{G.W. Gibbons and S.W. Hawking, Phys. Lett.  {\bf 78B} (1978) 430.}
\REF\gwg{G.W. Gibbons, Nucl. Phys. {\bf B207} (1982) 337.}
\REF\cmh{C.M. Hull, Phys. Lett. {\bf B139} (1984) 39.}
\REF\bgprt{E. Bergshoeff, M. de Roo, M.B. Green, 
G. Papadopoulos and P.K.
Townsend, Nucl. Phys. {\bf B470} (1996) 113.}
\REF\rruis{A. Dabholkar, G.W. Gibbons, 
J. Harvey and F. Ruiz Ruiz,  Nucl.
Phys. {\bf B340} (1990) 33; A. Dabholkar and 
J. Harvey, Phys. Rev.
Lett. {\bf 63} (1989) 478.}
\REF\paptown{G. Papadopoulos and P.K. Townsend, Phys. Lett.  {\bf 380B}
(1996) 273.}
\REF\ark{A. Tseytlin, Nucl. Phys. {\bf B475} (1996) 149.}
\REF\kleb{I. Klebanov and  A. Tseytlin, Nucl. Phys. {\bf B475} (1996) 179.}
\REF\jer{J.P. Gauntlett, D.A. Kastor and
 J. Traschen, Nucl. Phys. {\bf B478} (1996)
544.}
\REF\BBJ{K. Behrndt, E. Bergshoeff and B. Janssen, Phys. Rev. {\bf D55}
(1997) 3785, hep-th/9604168.}
\REF\costa{M.S. Costa, Nucl. Phys. {\bf B490} (1997) 202, hep-th/9609181.}
\REF\bmm{J.C. Breckenridge, 
G. Michaud and R.C. Myers, Phys. Rev. {\bf D55}
(1997) 6438, hep-th/9611174.}


\Pubnum{ \vbox{ \hbox{R/96/60} } }
\pubtype{Revised, May 15 and
 July 10}
\date{December, 1996}

\titlepage

\title{Superstring dualities and p-brane bound states}
\author{M.S. Costa and G. Papadopoulos}
\address{D.A.M.T.P
 \break University of Cambridge\break
         Silver Street \break Cambridge CB3 9EW}


\abstract {We show that the M-theory/IIA and
 IIA/IIB superstring  dualities
together with the diffeomorphism invariance 
of the underlying theories
require the presence of certain p-brane
 bound states in IIA and IIB
superstring theories preserving 1/2 of 
the spacetime supersymmetry.  We then
confirm the existence  of IIA and IIB 
supergravity solutions having the
appropriate p-brane bound states interpretation.}
\vskip 1cm


\endpage

\pagenumber=2


\def\cM {{\cal{M}}}



\chapter{Introduction}

Much evidence have been gathered in the past year to
 support the 
conjecture that  the strong coupling limit of 
IIA superstring theory is an
11-dimensional supersymmetric theory called
 M-theory [\pkta, \wita].  More
precisely,  the IIA superstring theory is
 the compactification of M-theory
on a $S^1$ with the IIA string coupling
 constant $\lambda$ related to the
compactification radius $R$ as 
$\lambda=R^{2/3}$. This interpretation
 of the IIA superstring theory 
requires the presence of non-perturbative
 states in IIA some of which should
carry Ramond-Ramond (RR) charges. The first
 indication that such states were
present in IIA superstrings was the existence
 of solutions of D=10 IIA
supergravity carrying the appropriate 
charges [\hs, \rrrbranes],
subsequently this was confirmed within
 IIA superstring theory by identifying
the IIA D-p-branes for p=0,2,4,6,8 as the
 carriers of RR charges [\pol].  
The conjectured D=11 Kaluza-Klein (KK) 
origin of IIA superstrings further
requires that {\it all} IIA p-branes  
should have a D=11 interpretation
[\pkta]. It turns out that this is the
 case with the IIA membrane and
5-brane being the `direct' reduction 
of M-branes, and  the IIA string and
4-brane being the `double' reduction 
of M-branes.   The rest of the IIA
p-branes, the D-0-branes and  
D-6-branes\foot{The IIA superstring theory
also has an 8-brane but no direct D=11 interpretation
 has been found for it
as yet.}, are interpreted as the KK modes and 
the KK monopoles of the
reduction from D=11 to D=10, respectively.  
In addition, the interpretation
of the IIA string and 4-brane as the double
 reduction of M-branes requires 
the existence of  BPS-saturated
 0-brane/p-brane bound states for p=1,4,
preserving precisely 1/4 of the spacetime
 supersymmetry [\doug,\gppkt]. The
corresponding supergravity solutions
 were given in [\gppkt] thus providing
more evidence for M-theory/IIA duality.

Another novel equivalence is  the  T-duality
 of IIA and IIB superstring 
theories. The IIA superstring theory 
compactified  on a circle of radius
$R_A$ is equivalent to the IIB superstring 
theory compactified on a circle
of radius  $R_B$ with $R_B=\alpha'/R_A$ where
$\alpha'$ is the string tension [\dhs, \dlp]. 
T-duality is a perturbative 
superstring symmetry and therefore this
 equivalence between IIA and IIB
superstrings  can be verified to all
 orders in string perturbation theory.  
As in the M-theory/IIA duality, the IIA
 p-branes transform under IIA/IIB
T-duality to the IIB p-branes and 
vice-versa. In particular, IIA D-p-branes
for p=0,2,4,6,8  transform under T-duality 
 either to the IIB D-(p-1)-branes
or to the IIB D-(p+1)-branes depending on 
whether the T-duality operation is
taken along a {\it worldvolume} or a {\it transverse}
 direction of the IIA
D-p-branes, respectively. Similarly,  the
 IIB D-p-branes for p=1,3,5,7,
transform under T-duality  either to the
 IIA D-(p-1)-branes or to the IIA
D-(p+1)-branes\foot{See  [\polch] for the T-duality 
transformation on the
D-branes from the superstring point of view.}.  
Furthermore in the
effective theory, the fundamental IIA (IIB) string
 and the solitonic IIA
(IIB) 5-brane transform under T-duality along a 
transverse direction to the
IIB (IIA) fundamental string and to the IIB (IIA)
 KK monopole, respectively. 
Moreover, the fundamental IIA  (IIB) string and 
the solitonic IIA (IIB)
5-brane transform under T-duality along a worldvolume
 direction  to the IIB
(IIA) plane wave and to the IIB (IIA) solitonic 
5-brane, respectively.

The KK reduction from M-theory to IIA superstring theory
 and the T-duality
operation from IIA (IIB) to IIB (IIA) require
 a choice of a spacetime 
direction.  But in diffeomorphic invariant
 theories, like M-theory, IIA and
IIB superstrings, there is no such prefered
 direction and therefore the
M-theory/IIA and IIA/IIB  dualities should be
 independent of this choice. 
For example, the M-theory should reduce to 
the IIA superstring along any
spatial D=11 direction.
However as we have  mentioned above, the
 interpretation
of p-branes after a reduction or a T-duality
 transformation 
depends on whether these operations are performed
  along  one of their
worldvolume or along one of their transverse
 directions. As we shall see,
a reduction or a T-duality transformation
 along a generic direction requires
the inclusion of new BPS-saturated states in 
the spectrum of IIA and IIB
superstrings. These BPS-saturated states are 
below threshold and  have the
interpretation of p-brane bound states
 preserving 1/2 of the spacetime
supersymmetry; we shall list these p-brane
 bound states in section 2.  The
presence of these p-brane bound states in IIA
 and IIB superstring theory is
{\sl required} by the M-theory/IIA
 and IIA/IIB  {\sl dualities} together
with the {\sl diffeomorphism invariance} of 
the underlying theories. This 
association of
 p-brane bound states to the  M-theory/IIA
 and IIA/IIB dualities is 
similar to the association of KK modes to
 the KK reduction.  Therefore, the
existence of these p-brane bound states in
 IIA and IIB superstrings is {\sl
necessary} for the consistency of these duality
 conjectures.  Some of the
p-brane bound states that we shall consider in 
this paper have already been
investigated either in the context of D-branes
 [\douglas,\polch,\gg,\sen,
\gilad] or from the effective theory point
 of view [\memdy,\gp,\glpt,\tsey].
They are also  required for the consistency
 of  other superstring duality
conjectures  [\swh,\wittenb].

We shall describe the above p-brane bound states
  from the macroscopic 
point of view.  However in certain cases
we shall also comment on their
microscopic properties. The evidence that
 we shall present for the existence
of these p-brane bound states and for
 their interpretation is derived from
consideration of the solutions of the
 associated effective supergravity
theories with the analogous interpretation.
 To derive the solutions in IIA
and IIB supergravity with the desirable
 interpretation, we shall start from
the solution of D=11 supergravity with
 the interpretation 
 of a membrane/fivebrane  bound state
  preserving 1/2 of the spacetime 
supersymmetry [\memdy]. This solution
 will be reduced to  solutions of IIA
supergravity using the `standard' reduction
 along either a {\sl worldvolume}
or a {\sl transverse} direction of the
 configuration [\gp,\glpt].  The
reduced solutions have a IIA interpretation
 as  bound states of two IIA
p-branes preserving 1/2 of the spacetime
 supersymmetry. Then starting from
these IIA solutions and using the T-duality
 rules [\bho], we shall construct
`standard'  T-duality chains as in [\gppkt], 
i.e. T-duality transformations
along {\sl worldvolume} or  {\sl transverse} 
directions of these
configurations, to find most of the required
 p-brane bound state solutions
in IIA and IIB supergravity theories.  These
  p-brane bound states will also
be derived using KK reductions and T-duality
  transformations from the 
p-brane solutions of D=11, IIA and IIB
 supergravities but this time these
operations will be taken along a generic
 direction in spacetime. In this
way, we shall establish that the origin
 of these bound states is due to the
M-theory/IIA  and IIA/IIB dualities and 
the diffeomorphism invariance of the
underlying theories.

The organization of this paper is as follows: 
In section two, we shall 
present the bound states of IIA and IIB 
superstring theories expected  from
the reduction of M-theory to IIA and from
 the IIA/IIB T-duality. In section
three, we shall give the IIA supergravity 
solutions with the interpretation of p-brane
bound states that are expected from
 the reduction of M-theory to IIA 
superstring.  In section four, we shall
 give the IIA and IIB supergravity
solutions with the interpretation of
 p-brane bound states that are expected
from the IIA/IIB duality. In section five, 
we shall present an M-theory
interpretation of some of these solutions, 
and in section six we shall give
our conclusions.


\chapter {p-brane bound states}

The reduction and T-duality operations do
 not commute with spacetime
diffeomorphisms.  Because of this there
 are many ways to choose a 
direction in spacetime to perform a 
reduction or a T-duality transformation.
However, as in [\tsey], here we shall
 consider  only  the
following two cases:  (i) In the first
 case, we shall perform the reduction
and the T-duality transformation along
 a spatial spacetime direction which
can be decomposed as a linear combination
 of a spatial worldvolume and a
transverse direction  of a p-brane, 
i.e. this direction intersects the
p-brane at an angle
$\alpha$.  We shall denote the corresponding
 operations with  $R_\alpha$ and
$T_\alpha$, and we shall refer to them as
 {\sl reduction at an angle} and
{\sl T-duality at an angle}, respectively. For
$\alpha=0$ or
$\alpha=\pi/2$,
$R_\alpha$ and $T_\alpha$ become the
 standard reduction and T-duality 
transformations along a worldvolume or 
a transverse directions of the
p-brane, respectively. Therefore
$R_\alpha$ ($T_\alpha$) is a linear 
combination of a standard reduction (T-duality
transformation) along a worldvolume
 direction and a standard reduction 
(T-duality transformation) along a 
transverse direction of a p-brane. (ii)
In the second case, we shall perform 
the reduction and the T-duality
transformation along a direction 
which is {\sl transverse} to a p-brane and
with the p-brane  moving with relativistic
 velocity $v$ in this direction,
i.e. the p-brane is `boosted' in this
 direction.  We shall denote the
corresponding operations with $R_v$ and
$T_v$, and we shall refer to them as
 {\sl reduction along a boost} and {\sl T-duality along a
boost}, respectively.  In what follows, 
we shall use
$R$ and $T$ to denote the `standard' reduction
 and the `standard' T-duality transformation
along either a {\it worldvolume} or a {\it transverse}
 direction of a 
p-brane.  We shall also use, when it is 
necessary in order to avoid
ambiguities, the subscripts $M, A$ and
$B$ to denote the BPS states of the 
corresponding theories, and the
subscripts $F,D$ and $S$ to denote 
the fundamental, Dirichlet and solitonic
p-branes, respectively.

As we have already mentioned in the introduction, 
reducing the M-branes 
along one of their worldvolume directions
 (double reduction) as
$$\eqalign{
2_M&{\buildrel R\over\longrightarrow}1_F
\cr
5_M&{\buildrel R \over\longrightarrow }4_D }
\eqn\redone
$$
 yields  the IIA fundamental string and D-4-brane,
while reducing them along one of their transverse directions (direct 
reduction) as
$$\eqalign{
2_M&{\buildrel R\over\longrightarrow}2_D
\cr
5_M&{\buildrel R \over\longrightarrow }5_S \ ,}
\eqn\redtwo
$$
 yields the IIA D-2-brane and the solitonic 5-brane. 
 Similarly, the
reductions of the (purely gravitational) D=11
 plane wave and the D=11 KK
monopole to D=10 are
$$\eqalign{
0_w&{\buildrel R\over\longrightarrow}0_w
\cr
0_w&{\buildrel R \over\longrightarrow }0_D \ ,}
\eqn\redthree
$$
and
$$\eqalign{
0_m&{\buildrel R\over\longrightarrow}0_m
\cr
0_m&{\buildrel R \over\longrightarrow }6_D \ ,}
\eqn\redfour
$$
respectively, where $0_w$ denotes the plane wave
 and $0_m$ denotes the
KK monopole. 

Now since $R_\alpha$ is equivalent to a simultaneous 
double and direct reduction,  reducing the
M-p-branes at an angle  
leads to IIA (p-1)-brane/p-brane bound states. 
It is convenient to use the
notation
$(r|p,q)$ to denote a solution  representing an r-brane
 intersection of a p-brane with a
q-brane; in the special case where
$p=r$, $(p|p,q)$ denotes a p-brane within a q-brane
 representing a  bound
state of a p-brane with a q-brane.  In this notation, 
the reduction of
M-branes at an angle is described as
$$\eqalign{
2_M&{\buildrel R_\alpha\over\longrightarrow}(1|1,2)_A
\cr
5_M&{\buildrel R_\alpha \over\longrightarrow }(4|4,5)_A \ .}
\eqn\pone
$$
Similarly, the reduction of the D=11 plane wave and 
the D=11 KK  monopole at
an angle leads to the IIA bound states
$$
\eqalign{
0_w&{\buildrel R_\alpha\over\longrightarrow }(0_w|0)_A
\cr
0_m&{\buildrel R_\alpha\over\longrightarrow }(0_m|6)_A\ ,}
\eqn\ptwo
$$
where $(0_w|0)_A$ denotes a bound state of a 
plane wave with a 0-brane, and
$(0_m|6)_A$ denotes the bound state of a 
KK monopole with a 6-brane.  The
$(0_w|0)_A$ bound state is a `boosted' 0-brane. 

Next, the reduction of the M-branes, 
the D=11 plane wave and the D=11 KK 
monopole along a boost
 leads to IIA bound states that
always involve the IIA D-0-brane.  This 
is because from the D=10 
perspective the (quantised) momentum in the
 compactifying direction becomes
the mass of the KK modes of the compactification. 
But  as we have mentioned
in the introduction, these are identified 
with the IIA D-0-branes. So this
observation together with the fact that
 we reduce along a transverse
direction of the configurations lead to
 the following  IIA p-brane bound
states: 
$$
\eqalign{
2_M&{\buildrel R_v\over\longrightarrow }(0|0,2)_A
\cr
5_M&{\buildrel R_v\over\longrightarrow }(0|0,5)_A
\cr
0_w&{\buildrel R_v\over\longrightarrow }(0|0_w)_A
\cr
0_m&{\buildrel R_v\over\longrightarrow }(0|0,6)_A \ .}
\eqn\pthree
$$
We remark that it has been argued in [\polch]
 that there is not a IIA  BPS 
D-0-brane/D-6-brane bound state. We shall
 return to this point when we
discuss the $(0|0,6)_A$ solution of IIA
 supergravity in the next section.
As in \ptwo, the $(0_w|0)_A$
 bound state is a `boosted' 0-brane.

We now turn to the p-brane bound states
 required by the  IIA/IIB superstring
duality.  As we have already mentioned
 T-duality  transforms a 
(IIA or IIB) D-(p+1)-brane  either as
$$
(p+1)_D{\buildrel T\over\longrightarrow } p_D
\eqn\tdone
$$
 or as
$$
(p+1)_D{\buildrel T\over\longrightarrow } (p+2)_D
\eqn\tdtwo
$$
depending on whether the T-duality 
transformation is performed along a
worldvolume or a transverse direction 
of the D-(p+1)-brane, respectively, 
where the p-brane and the (p+2)-brane
 are (IIB or IIA) D-branes. Similarly,
the IIA or IIB fundamental string, 
solitonic 5-brane, plane wave and KK
monopole transform under T-duality as follows:
$$
\eqalign{
1_F&{\buildrel T\over\longleftrightarrow } 1_F\ , \qquad\qquad
1_F{\buildrel T\over\longleftrightarrow } 0_w
\cr
5_S&{\buildrel T\over\longleftrightarrow } 5_S
\ , \qquad\qquad
5_S{\buildrel T\over\longleftrightarrow } 0_m
\cr
0_w&{\buildrel T\over\longleftrightarrow } 0_w
\ , \qquad\qquad
0_m{\buildrel T\over\longleftrightarrow } 0_m\ .}
\eqn\tdfour
$$
We remark that the transformation 
$1_F\leftrightarrow 0_w$ is the usual 
exchange between (IIA or IIB) winding
 modes and (IIB or IIA) momentum
modes of the fundamental string under T-duality.

Now since $T_\alpha$ is a linear combination
 of a T-duality transformation 
along a worldvolume direction and a T-duality 
transformation along a
transverse direction of a p-brane, it is clear
 from \tdone\ and \tdtwo\ that
acting with a T-duality at an angle on a 
(IIA or IIB) D-(p+1)-brane will
lead to  BPS-saturated (IIB or IIA)
 D-p-brane/D-(p+2)-brane bound states
preserving 1/2 of the spacetime supersymmetry, i.e.
$$
(p+1)_D{\buildrel T_\alpha\over\longrightarrow }
 \big(p|p_D,(p+2)_D\big)\ .
\eqn\pfour
$$
The  $\big(p|p_D,(p+2)_D\big)$, for $p=0,2,4,6$,
 are bound states of IIA 
superstring theory and  $\big(p|p_D,(p+2)_D\big)$, 
for $p=1,3,5$, are bound
states of IIB superstring theory. 
The existence  of the  bound states 
$(p|p_D,(p+2)_D)$ preserving 1/2 of 
the spacetime supersymmetry is also 
expected from D-brane considerations.  
Similarly, applying T-duality at an 
angle to the (IIA or IIB)
fundamental string,  solitonic 5-brane, 
plane wave and KK monopole leads 
to the following bound states:
$$
\eqalign{
1_F&{\buildrel T_\alpha\over\longrightarrow }(0_w|1_F)
\cr
5_S&{\buildrel T_\alpha\over\longrightarrow }(0_m|5_S)
\cr
0_w&{\buildrel T_\alpha\over\longrightarrow }(0_w|1_F)
\cr
0_m&{\buildrel T_\alpha\over\longrightarrow }(0_m|5_S)\ .}
\eqn\pfive
$$
The bound state $(0_w|1_F)$ is simply a boosted fundamental string.

Next, the p-brane bound states expected
 from applying  T-duality along a 
boost to the IIA and IIB p-branes  will
  always involve the fundamental (IIA
or IIB) string.   This is because the 
(quantised) momentum modes along the
direction of the T-duality transformation become
 in the dual picture the
winding modes of the (IIA or IIB) {
\it fundamental} string.  This observation together
 with the fact that
the direction of the boost is transverse to 
the objects lead to the 
following bound states involving the
 (IIA or IIB) D-p-branes:
$$
p_D{\buildrel T_v\over\longrightarrow } \big(1|1_F,(p+1)_D\big)
\eqn\psix
$$
where $p=0,...,7$.   An alternative explanation
 for the presence  of  the
fundamental string/D-(p+1)-brane bound state
 can be found from consideration
of the D-p-brane effective action. In this case, 
T-duality turns velocity in
a transverse direction of a p-brane to an 
electromagnetic field in the dual
picture.  But the flux of the electomagnetic
 field is related to the tension
of the fundamental string [\pktb], so in 
the dual picture one finds a
fundamental string/D-(p+1)-bound state.    
 Similarly, the expected bound
states from applying T-duality along a 
boost to the (IIA or  IIB) fundamental
string, solitonic 5-brane, plane wave and KK monopole  are
$$
\eqalign{
1_F&{\buildrel T_v\over\longrightarrow }1_F
\cr
5_S&{\buildrel T_v\over\longrightarrow }(0_m|1_F)
\cr
0_w&{\buildrel T_v\over\longrightarrow }(0_w|1_F)
\cr
0_m&{\buildrel T_v\over\longrightarrow }(1|1_F,5_S)\ .}
\eqn\pfivea
$$
We remark that, as we shall see in section 4,  
the supergravity solution
associated with the $(1|1_F,5_S)$ 
bound state has similar qualitative
features as the solution that is
 interpreted as a $(0|0,6)_A$ bound state.

So far we have listed the bound states
 that we expect to find in IIA and 
IIB superstring theories due to the
 M-theory/IIA and IIA/IIB dualities and
to the diffeomorphism invariance of 
the underlying theories. Evidence for
the existence of all these bound states
 will be given in the next two
sections by finding the solutions of
 the corresponding supergravity theories
with the appropriate interpretation. 
We have been able to carry out this
computation for all the required 
p-brane bound states. The  resulting IIA
and IIB p-brane bound state solutions 
will be expressed in the {\sl string}
frame.  All of them have a parameter 
$\alpha$ which interpolates between the
two constituent objects that form the  bound states.


\chapter{M-theory/IIA duality and bound states}

 As we have explained in the
previous section, M-theory/IIA
 duality  requires  the existence of
the bound states $(1|1,2)_A$ and $(4|4,5)_A$ 
in IIA superstring theory
preserving 1/2 of the spacetime supersymmetry. 
The corresponding
supergravity  solutions can be obtained starting
 from the D=11 supergravity
solution $(2|2,5)_M$ [\memdy], 
$$
\eqalign{
ds^2&=(H \tilde H)^{{1\over 3}}\big[H^{-1} ds^2(\bM^3)+ \tilde
H^{-1}ds^2(\bE^3)+ds^2(\bE^5)\big]
\cr
G_4&=\sin\alpha\, \epsilon(\bM^3)\wedge dH^{-1}-\cos\alpha\,
\star dH +\tan\alpha\,
\epsilon(\bE^3)\wedge d\tilde H^{-1}\ ,}
\eqn\asolone
$$
preserving 1/2 of the spacetime supersymmetry, 
where $H$ is a harmonic
function on $\bE^5$, $\tilde H=
\sin^2\alpha + \cos^2\alpha\, H$, star  is
the Hodge star in $\bE^5$ and
$\alpha\in [0,\pi/2]$ is an angle parameter
 of the solution; this  solution
interpolates between the membrane and 
fivebrane solutions of D=11
supergravity theory by adjusting the parameter
$\alpha$. In what follows, we shall take 
the harmonic function $H\sim 1$  at
the transverse spatial infinity. The IIA
$(1|1_F,2)_A$ and
$(4|4,5_S)_A$ solutions can be derived from
 the following T-duality chains: 
$$
(2|2,5)_M{\buildrel  R \over\rightarrow} 
(1|1_F,4)_A{\buildrel  T
\over\rightarrow}(1|1_F,3)_B 
{\buildrel  T \over\rightarrow} (1|1_F,2)_A
\eqn\solonea
$$
and
$$
(2|2,5)_M{\buildrel  R \over\rightarrow}
 (2|2,5_S)_A{\buildrel  T
\over\rightarrow}(3|3,5_S)_B 
{\buildrel  T \over\rightarrow} (4|4,5_S)_A\ .
\eqn\soloneb
$$

It should  also be possible to construct
 the  $(1|1,2)_A$ and
$(4|4,5)_A$ solutions of IIA supergravity
 by KK reduction of the  M-brane
solutions at an angle. This computation
 was done in [\tsey] by first
rotating  the M-brane solutions and 
then reducing them to D=10. Here we
shall repeat this computation and relate 
the result to that of the T-duality
chains \solonea\ and \soloneb. 
The M-brane solutions   [\duff,\guv] are 
$$
\eqalign{
ds^2&=H^{(p+1)/9}\big[H^{-1} ds^2(\bM^{p+1})+ds^2(\bE^{10-p})\big]
\cr
G_4&=\cases{\epsilon(\bM^3)\wedge dH^{-1}\ ,\qquad p=2 \cr
            -\star dH\ ,\qquad\qquad\quad p=5}}
\eqn\solone
$$
where $\bE^n$ is  the Euclidean space of
 dimension $n$, $\bM^n$ is   the
Minkowski  space of dimension
$n$, $H$ is a harmonic function on
 $\bE^{10-p}$, $\epsilon(\cM)$ is  the
volume   form of a manifold $\cM$, 
and the star is the Hodge star in
$\bE^5$.  We also write for later use 
 the (purely gravitational)  D=11
plane wave solution 
$$
ds^2=-dt^2+d\rho^2+(H-1) (dt+d\rho)^2+ds^2(\bE^9)\ ,
\eqn\ppwave
$$
where $H$ is a harmonic function on $\bE^9$, 
and the D=11 KK monopole solution 
$$
ds^2=ds^2(\bM^7)+ds^2(G/H)\ ,
\eqn\monopole
$$
where $ds^2(G/H)=H^{-1}(d\rho+\omega)^2+H ds^2(\bE^3)$ is the 
Gibbons-Hawking metric [\gh], $H$ is a
harmonic function on $\bE^3$ and
$d\omega={}^*dH$. Both the plane wave 
[\gwg, \cmh] and the KK monopole
preserve 1/2 of the spacetime supersymmetry; 
for the latter case this
follows from the fact that $ds^2(G/H)$ 
is hyper-K\"ahler.  To choose the
compactifying  direction
$u_{11}$ in D=11, we use a D=11  
diffeomorphism that mixes the  spatial
worldvolume and the transverse space
 coordinates of the M-branes.  An example
of such diffeomorphism is
$$
\pmatrix{u_{10}\cr u_{11}}=A \pmatrix{\rho\cr y}
\eqn\soltwo
$$
where $\rho$ is the coordinate along a 
spatial worldvolume direction, $y$ is 
the coordinate
along a transverse direction and $A$ is 
a real invertible $2\times 2$ matrix. Let us
choose 
$$
A=\pmatrix { \cos\alpha&\quad  -
\sin\alpha\cr \sin\alpha& \quad \cos\alpha}
\eqn\asoltwo
$$ 
 where $\alpha$ is the angle that appears 
as a parameter in the $(2|2,5)_M$
solution above.  The solutions of IIA 
supergravity that are constructed  by
reducing the M-branes  along
$u_{11}$ [\tsey] are identical 
to the solutions found by the duality 
chains \solonea\ and
\soloneb\ applied to the $(2|2,5)_M$ solution. 
This confirms that the 
presence of
$(1|1,2)_A$ and
$(4|4,5)_A$ solutions is a consequence 
of the M-theory/IIA  duality and
the diffeomorphism invariance of M-theory.  
Conversely, the existence  of
these solutions serve as further 
evidence for the M-theory/IIA  duality. 

 As we have already mentioned in section 2, 
reducing the D=11 plane  wave
and the KK monopole at an angle leads 
to the bound states
$(0_w|0)_A$ and 
$(0_m|6)_A$, respectively. The 
$(0_w|0)_A$ solution may be found by 
extending the T-duality chain
\solonea\ as
$$
(1|1_F,2)_A{\buildrel  T
\over\rightarrow}(1|1_F,1_D)_B 
{\buildrel  T \over\rightarrow}(0_w|0)_A\ ,
\eqn\soltwoa
$$
where  $(0_w|0)_A$ is
$$
\eqalign{
ds^2&=\tilde H^{{1\over2}}\big [-H^{-1} dt^2+ H \tilde H^{-1} (dx+
\sin\alpha\, (H-1) H^{-1}dt)^2+ ds^2(\bE^8)\big ]
\cr
e^\phi&=\tilde H^{{3\over4}}
\cr
F_2&=\cos^{-1}\alpha\, dt\wedge d\tilde H^{-1}+\tan\alpha\, dx\wedge
d\tilde H^{-1}\ .}
\eqn\asoltwoa
$$
This solution is interpreted as a 
`boosted' 0-brane and  is  diffeomorphic
(up to a rescalling of its mass, 
i.e. a rescalling of ($H-1$)) to the
0-brane solution of IIA supergravity.  
Note that varying the parameter
$\alpha$, the solution \asoltwoa\ 
interpolates between the IIA plane wave
solution ($\alpha=\pi/2$)  and the 
IIA D-0-brane ($\alpha=0$). 
 Similarly, the 
$(0_m|6)_A$ solution  can be obtained 
by extending the T-duality  chain
\soloneb\ as
$$
(4|4,5_S)_A{\buildrel  T
\over\rightarrow}(5|5_D,5_S)_B 
{\buildrel  T \over\rightarrow} (0_m|6)_A\ ;
\eqn\soltwob
$$
the $(0_m|6)_A$ solution is
$$
\eqalign{ds^2&=\big(\tilde H H\big)^{{1\over2}}
\big[ H^{-1}ds^2(\bM^6)
+\big(\tilde H H\big)^{-1}(dx+ \cos\alpha\, \omega)^2+ ds^2(\bE^3) \big]
\cr
e^\phi&=\big(\tilde H H^{-1}\big)^{{3\over4}}
\cr
F_2&=- \sin\alpha\, d(\omega \tilde H^{-1})+\tan\alpha\, dx\wedge
d\tilde H^{-1}\ ,}
\eqn\solthree
$$
where $d\omega={}^*dH$ and  the 
Hodge duality is with respect to the 
Euclidean metric in $\bE^3$. Both 
the $(0_w|0)_A$ and the $(0_m|6)_A$
solutions can also be obtained from 
the D=11 plane wave and the KK monopole
by first rotating the solutions using
 \soltwo\ and \asoltwo, and then
reducing them to D=10 along $u_{11}$. 
 For the $(0_w|0)_A$ solution this was
done in [\tsey].

Next we turn to the IIA supergravity solutions
 associated with the  bound
states $(0|0,2)_A$  and
$(0|0,5)_A$. These solutions can be easily
 derived from the T-duality chains 
$$
(2|2,5)_M{\buildrel  R \over\rightarrow} 
(2|2,4)_A {\buildrel  T 
\over\rightarrow} (1|1_D,3)_B 
{\buildrel  T \over\rightarrow} (0|0,2)_A
\eqn\solbthree
$$
and
$$
(2|2,5)_M{\buildrel  R \over\rightarrow} 
(2|2,5)_A {\buildrel  T 
\over\rightarrow} (1|1_D,5_S)_B 
{\buildrel  T \over\rightarrow} (0|0,5)_A\ .
\eqn\solbfour
$$
The $(0|0,2)_A$ and $(0|0,5)_A$ solutions
 are also obtained by  reducing the
M-brane solutions along a boost with 
velocity $v/c=\sin\alpha$ in the
direction of compactification, where 
$c$ is the speed of light 
(in our units $c=1)$)[\tsey].    
The  appropriate
D=11 diffeomorphism which relates 
the compactifying coordinate
$u_{11}$ to the
 worldvolume time-coordinate $t$ and
 transverse coordinate $y$ of  the
M-branes is
$$
u_{11}=\cos^{-1} \alpha\,\big(y+\sin\alpha\, t\big), \qquad t'= 
\cos^{-1}\alpha\,
\big(t+\sin\alpha\, y\big)\ .
\eqn\solbone
$$
We also have to  `boost' the 
harmonic function $H$ (i.e. the mass) of the
M-branes as 
$$
(1-H)\rightarrow \cos^2\alpha\, (1-H)\ .
\eqn\solbtwo
$$
Similarly, reducing the D=11 plane 
wave along a boost we get the 
$(0|0_w)_A$  solution (eqn.
\asoltwoa).  Finally, boosting the D=11
KK monopole, eqn.\monopole, along $\rho$  using
\solbone\ (setting $y=\rho$) and
 then reducing along $u_{11}$, we get the  
IIA supergravity solution
$$
\eqalign{ds^2&=\big(\bar H \tilde H\big)^{{1\over2}}
\big[- (\bar H \tilde
H)^{-1}(dt+
\sin\alpha \cos\alpha\,
\omega)^2+\tilde H^{-1} ds^2(\bE^6)+ds^2(\bE^3) \big] 
\cr
e^\phi&=(\bar H \tilde H^{-1}\big)^{{3\over4}}
\cr
F_2&=\sin^{-1}\alpha\, dt\wedge d\bar H^{-1}- \cos\alpha\, d(\omega \bar
H^{-1})\ , }
\eqn\asolbtwo
$$
where
$$
\bar H=1+\sin^2\alpha\,(1-H)\ ,
\eqn\absolbtwo
$$ 
and $d\omega={}^*dH$; the Hodge duality
 is with respect to the  flat metric
in $\bE^3$.  We have {\sl not} been
able to derive this solution using a
 T-duality chain from the D=11
$(2|2,5)_M$ solution as we have done 
for the rest of the p-brane  bound state
solutions.  Observe that this metric 
has an off-diagonal term which is
reminiscent to a similar term in the 
Taub-NUT metric. This solution
appears to have the appropriate 
non-vanishing supergravity field strength
to be interpreted as a 0-brane/6-brane bound state. 
 However although for
$\alpha=0$ we find the familiar 6-brane solution
 of IIA supergravity, for
$\alpha={\pi\over2}$ we find the 0-brane solution
 but with {\it negative} ADM
mass.   In addition, the solution, apart from 
the singularities at the centres of the harmonic
function $H$,  has another singularity at
$\bar H(r_0)=0$ which occurs at small 
string coupling whenever $\sin\alpha\not=0$. 
Note that in order to boost the KK
monopole one needs to
\lq decompactify' the coordinate $\rho$ along 
the killing direction in order to impose the
periodicity in the compactifying
$u_{11}$ coordinate. This re-introduces the
 NUT singularities of the KK monopole that were
resolved by the periodic identification of
$\rho$. It is not clear though that there is a 
direct relation between the NUT
singularities of the KK monopole and the 
$\bar H(r_0)=0$ ones since in any case 
they occur at
different points. It is more likely that 
the NUT singularities of the KK monopole are
related to the singularities of the D=10 
solution that occur at the centres of $H$.  
The above properties of the solution 
obscure its interpretation as a
0-brane/6-brane bound state.    This 
appears to be in
agreement with the result of [\polch] that
 there does not exist such a BPS bound state.   

\chapter{IIA/IIB T-duality and bound states}

As we have explained in the section 2, the
IIA/IIB superstring T-duality and the
 diffeomorphism invariance of  the
underlying theories requires the existence
 of D-p-brane/D-(p+2)-brane bound
states in IIA and IIB superstring theories.  
The D-p-brane bound states
present  in IIA superstring theory  are
$(0|0,2)_A$,
$(2|2,4)_A$, $(4|4,6)_A$ and
$(6|6,8)_A$, and those present in IIB  
are $(1|1_D,3)_B$, $(3|3,5_D)_B$ and
$(5|5_D,7)_B$.  
To find the corresponding IIA supergravity solutions,
 recall that the $(2|2,4)_A$ solution can 
be obtained from the  $(2|2,5)_M$
solution of M-theory by reducing along one 
of the fivebrane worldvolume
directions orthogonal to the membrane 
[\gp,\glpt]. Having found $(2|2,4)_A$,
we then construct a T-duality chain that
 extends to the `left' and the
`right' of this solution as
$$
(0|0,2)_A{\buildrel  T \over\rightarrow}
(1|1_D,3)_B {\buildrel  T
\over\rightarrow}(2|2,4)_A{\buildrel  T
\over\rightarrow}(3|3,5_D)_B
{\buildrel  T \over\rightarrow}(4|4,6)_A{\buildrel  T
\over\rightarrow}(5|5_D,7)_B{\buildrel  T
\over\rightarrow}(6|6,8)_A\ ,
\eqn\abone
$$
where the 7-brane is the circularly 
symmetric one  and in last
step of this T-duality chain we have
 used the `massive' T-duality  rules of
[\bgprt]. The solutions representing 
the D-p-brane/D-(p+2)-brane bound
states can be easily computed from the
 $(2|2,4)_A$ solution using the chain
\abone. These D-p-brane/D-(p+2)-brane 
 bound states solutions should also be
derivable using  T-duality transformations 
at an angle on the  IIA and IIB
D-(p+1)-branes, i.e. applying a T-duality 
transformation along the direction
$u_{11}$ of
\soltwo. It turns out that this is the case
 and the solutions obtained 
from the two different ways of doing the 
computation coincide.
 As an example we derive the IIA $(0|0,2)_A$ 
bound state solution from  the
IIB D-1-brane solution. The D-1-brane solution is 
$$
\eqalign{
ds^2&=H^{-{1\over2}} ds^2(\bM^2)+ H^{{1\over2}}ds^2(\bE^8)
 \cr
F^{(2)}_3&=-\epsilon(\bM^2)\wedge dH^{-1}
\cr
e^\varphi&=H^{{1\over2}} \ ,}
\eqn\solfour
$$
where $\varphi$ is the IIB dilaton. We 
parameterise $\bM^2$ with the  
coordinates
$(t,\rho)$, and $\bE^8$ with the coordinates 
$\{y^1,\dots, y^8\}$ and  set
$y=y^1$ . Then we change coordinates 
from $(\rho,y)$ to $(u_{10},u_{11})$ as
in \soltwo\ and \asoltwo, and  perform a 
T-duality transformation along
$u_{11}$.  The resulting IIA $(0|0,2)_A$ solution is
$$
\eqalign{
ds^2&=-H^{-{1\over2}}dt^2+H^{{1\over2}} 
\tilde H^{-1}ds^2(\bE^2)+ 
H^{{1\over2}}ds^2(\bE^7)
\cr
e^\phi&=H^{{3\over4}} \tilde H^{-{1\over2}}
\cr
F_2&=-\sin\alpha\, dt\wedge dH^{-1} 
\cr
F_4&=\cos\alpha\, \epsilon(\bM^3) \wedge dH^{-1}
\cr
F_3&=\tan\alpha\, \epsilon(\bE^2)\wedge d\tilde H^{-1}\ ,}
\eqn\solfive
$$
in agreement with [\tsey], 
where $\phi$ is the IIA dilaton.  
Similarly, the bound states $(0_w|1_F)$ 
and $(0_m|5_S)$  expected   from
applying T-duality at an angle to the 
(IIA or IIB) fundamental string, the
solitonic 5-brane, the plane wave and 
the KK monopole are as follows:  The 
$(0_w|1_F)$ solution is
$$
\eqalign{
ds^2&=-H^{-1} dt^2+ H \tilde H^{-1} 
(dx+\sin\alpha\, (H-1) H^{-1}dt)^2+ 
\tilde H^{-1}dz^2+ds^2(\bE^7)
\cr
e^\phi&=\tilde H^{-{1\over2}}
\cr
F_3&=-\cos^{-1}\alpha\, dt\wedge dz\wedge d\tilde H^{-1}-\tan\alpha 
dx\wedge dz\wedge d\tilde H^{-1}\ ,}
\eqn\solsix
$$
and it can be interpreted as a boosted 
fundamental string so it is 
diffeomorphic to the fundamental string 
solution [\rruis] (up to a
rescalling of $H-1$).  The
$(0_m|5_S)$ solution  is
$$
\eqalign{
ds^2&=ds^2(\bM^5)+H \tilde H^{-1} dz^2+ 
\tilde H^{-1} (dx+\cos\alpha\,
\omega)^2+  H ds^2(\bE^3)
\cr
e^\phi&=H^{{1\over2}}\tilde H^{-{1\over2}}
\cr
F_3&=-\sin\alpha\, dz\wedge d(\omega 
\tilde H^{-1})-\tan\alpha\, dx\wedge
dz\wedge d\tilde H^{-1}\ ,}
\eqn\solsix
$$
where $\omega$ is as in \asolbtwo.

Next we turn to examine the solutions  
corresponding to the p-brane   bound
states which are  expected from 
applying a T-duality along a
boost to the  (IIA or IIB) D-p-branes. 
As we have explained in section 2,
these bound states always involve the 
(IIA or IIB) fundamental string.  The
corresponding solutions can be found 
from the T-duality chains
$$
(2|2,5)_M {\buildrel  R \over\rightarrow} 
(1|1_F,4)_A{\buildrel  T \over
\rightarrow} (1|1_F,3)_B {\buildrel  T
\over\rightarrow}(1|1_F,2)_A{\buildrel  T
\over\rightarrow}(1|1_F,1_D)_B\ ,
\eqn\solseven
$$
and 
$$
(2|2,5)_M {\buildrel  R \over\rightarrow}(1|1_F,4)_A
{\buildrel  T \over
\rightarrow} (1|1_F,5_D)_B {\buildrel  T
\over\rightarrow}(1|1_F,6)_A{\buildrel  T
\over\rightarrow}(1|1_F,7)_B {\buildrel  T
\over\rightarrow}(1|1_F,8)_A\ .
\eqn\soleight
$$
It turns out that these bound states can also 
be derived by first  using the
D=10 diffeomorphism \solbone\ together 
with \solbtwo, and
then performing a T-duality 
transformation along $u_{11}$.  Thus 
confirming our interpretation in 
section 2 for the origin of these bound
states. We shall not present all these 
solutions here since they can be
easily derived by applying the T-duality 
rules [\bho]. In any case, some of
these solutions are already known like 
for example the
$(1|1_F,4)_A$  and the $(1|1_F,1_D)_B$ solutions; 
the latter is associated 
with the IIB (p,q) strings[\swh, \wittenb]. 
Nevertheless, we present the
$(1|1_F,6)_A$ solution 
$$
\eqalign{ds^2&=\tilde H^{{1\over2}}
\big[H^{-1}ds^2(\bM^2)+\tilde H^{-1}
ds^2(\bE^5)+ds^2(\bE^3)\big]
\cr
e^\phi&=\tilde H^{-{1\over4}} H^{-{1\over2}}
\cr
F_3&=-\sin \alpha\, \epsilon(\bM^2)\wedge dH^{-1}+\sin\alpha\, 
\cos\alpha\, dt\wedge d(H^{-1}\omega)
\cr
F_2&=\cos \alpha \star dH }
\eqn\asoleight
$$
as another example, where $\omega$ is 
given as in \asolbtwo.
Next applying  T-duality  along 
a boost to the
(IIA or IIB) fundamental string , 
we do not find a new bound
state. To be more precise, the D=10 
solution that one finds can be  brought
into the form of that of the (IIB or IIA)
 fundamental string up to a D=10
diffeomorphism. Similarly, applying T-duality 
along a boost to the plane
wave we get the $(0_w|1_F)$ solution 
as in eqn. \solsix.  The
$(0_m|1_F)$ bound state solution 
associated with a boosted solitonic (IIA or
IIB) 5-brane is
$$
\eqalign{
ds^2&=-\tilde H H^{-1}dt^2+ds^2(\bE^5)+
H^{-1}(dx+\cos\alpha\, \omega)^2+
\tilde H ds^2(\bE^3)
\cr
F_3&=\sin\alpha\, dt\wedge dx\wedge 
dH^{-1}- \cos\alpha\, \sin\alpha\,  dt
\wedge d(\omega H^{-1})
\cr
e^\phi&=\tilde H^{{1\over2}} H^{-{1\over2}}\ , }
\eqn\solnine
$$
where $\omega$ is as in \asolbtwo. 
 Finally, applying T-duality along  a
boost to the KK monopole yields the
$(1|1_F,5_S)$ solution 
$$
\eqalign{ds^2&=-\bar H^{-1}(dt+
\sin\alpha\, \cos\alpha\,
\omega)^2+ds^2(\bE^5)+\bar H^{-1} 
\tilde H dx^2+\tilde H ds^2(\bE^3) 
\cr
e^\phi&=\big(\tilde H \bar H^{-1}\big)^{{1\over2}}
\cr
F_3&=\sin^{-1}\alpha\, dt\wedge dx
\wedge d\bar H^{-1}+ \cos\alpha\, dx\wedge
d(\omega
\bar H^{-1})\ , }
\eqn\solbten
$$
where $\bar H$ and $\omega$ is given 
as in \asolbtwo. The interpretation
of this solution as a fundamental 
string/solitonic 5-brane bound state
is obscured for similar reasons to that
 of the interpretation of the
solution \asolbtwo\ as a 0-brane/6-brane 
bound state. We shall not repeat the argument here.

\chapter{M-theory and p-brane bound states}

Some of the bound state solutions of 
IIA supergravity  derived in the 
previous two sections can be lifted  
to eleven dimensions to yield new BPS
solutions of D=11 supergravity preserving 1/2 of the spacetime
supersymmetry.  The bound states 
that we shall lift here are
$(0_m|5_S)$,
$(4|4,6)_A$,
$(1|1,6)_A$ and
$(0_m|1_F)$. 
The D=11 solution that yields $(0_m|5_S)$ or
 $(4|4,6)_A$ after reduction has the
interpretation of a D=11 fivebrane/KK monopole bound state and it can be
expressed as 
$$
\eqalign{ds^2=&H^{{2\over3}}
\tilde H^{{1\over3}}\big[ H^{-1} ds^2(\bM^5)+
\tilde H^{-1}ds^2(\bE^2) 
\cr &
+(H\tilde H)^{-1} \big(dx+\cos\alpha\, \omega \big)^2+ds^2(\bE^3)\big]
\cr
G_4=&-\sin\alpha\, \epsilon(\bE^2)\wedge d(\omega \tilde H^{-1})-\tan\alpha\,
\epsilon(\bE^2)\wedge dx\wedge d\tilde H^{-1}\ ,}
\eqn\mone
$$
where $\omega$ is as in \asolbtwo.
Finally the D=11 solution that yields
 $(1|1,6)_A$ or $(0_m|1_F)$  after 
reduction has the interpretation of 
 a D=11 membrane/KK monopole bound state
and it can be expressed as
$$
\eqalign{ds^2=&\tilde H^{{2\over3}}
 H^{{1\over3}}\big[- H^{-1} dt^2+\tilde
H^{-1} ds^2(\bE^5)+H^{-1} dz^2 
\cr &+(H\tilde H)^{-1} \big(dx+\cos\alpha\, 
\omega \big)^2+ds^2(\bE^3)\big]
\cr
G_4=&\sin\alpha\, \cos\alpha\, dt\wedge dz
\wedge d(\omega H^{-1})+\sin\alpha\, dt\wedge
dz\wedge dx\wedge d H^{-1}\ .}
\eqn\mtwo
$$

\chapter{Concluding remarks}

It is well known that apart from the
 M-brane solutions of D=11 supergravity, there are
other solutions that have the 
interpretation of intersecting M-branes 
[\paptown].  The standard  reduction of 
these configurations [\paptown,
\ark, \gp, \kleb, \jer, \BBJ] to D=10 has 
been extensively  studied yielding
solutions in D=10 with the interpretation of 
intersecting IIA p-branes.  It
is clear that one can reduce the 
intersecting M-brane solutions along a
generic D=11 direction
$u$.  There are many ways to choose such  
a D=11 direction.  For this,  let
us consider the intersecting M-brane solution
$(k|p_1,\dots, p_\ell)$.  Then $u$ can be 
chosen  as a linear combination 
of (i)  a k-brane spatial worldvolume 
direction and a `relative transverse'
direction of the  configuration, (ii)  
a k-brane spatial worldvolume
direction and a `overall transverse' 
direction of the configuration, (iii)
 two `relative transverse' directions of 
the configuration with the metric
having different  components along 
these directions and (iv)  a  `relative
transverse' and a `overall transverse' 
directions of the configuration. Each
such choice will lead to a different
 D=10 reduction of the D=11 solution. 
We can also  boost the intersecting 
M-brane solutions in different ways and
then reduce them   along the direction 
of the boost thus producing many more
new bound states in D=10.    The same  
applies for the T-duality
transformations acting on the IIA 
and IIB intersecting brane
configurations.  Thus many more 
solutions can be constructed in D=10 by
chains of T-duality transformations all  
preserving the same amount of
spacetime supersymmetry as the original solution.  
Most of these new
solutions have the interpretation of 
bound states of intersecting p-branes
or that of intersecting p-brane bound states [\costa]. 

The argument that we have used to 
propose the existence of p-brane bound 
states in IIA and IIB superstring theories
 also applies in $D<10$.  Apart
from the p-brane bound states that are 
obtained by reducing the D=10 p-brane
bound states that we have found in 
section 2 to lower dimensions, there
should also exist  additional p-brane 
bound states in $D<10$.  This will be
the case whenever a duality between 
two superstring theories breaks some of
their reparameterisation invariance.

\vskip 0.5cm
\noindent{ \sl Note added in the proof}
\vskip 0.05cm
During the preparation of this paper we 
have received  [\bmm] in which the
solutions interpreted as
 D-p-brane/D-(p+2)-brane bound states are 
constructed from the D-brane solutions
 using rotations and T-duality
transformations. This paper overlaps  
with some of our material in section 4.
\vskip 0.5cm

\noindent{\bf Acknowledgments:} We would like to 
thank M. Douglas,  G.W.
Gibbons, M.B. Green, C.M. Hull,  
M. Perry and P.K. Townsend for helpful
comments. G.P. is supported by a University 
Research Fellowship from the
Royal Society. M.S.C. is funded by 
JNICT (Portugal) under the programme
PRAXIS XXI.

\refout

\bye